\documentclass[epj-spec]{svjour}
\usepackage{graphicx}
\usepackage{amssymb}

\newcommand{ \rts }{$\sqrt{s_{_{\rm NN}}}$}

\newcommand{\nbin}{$N_{bin}$}

\newcommand{\raa}{$R_{AA}$}
\newcommand{\raajet}{$R_{AA}^{Jet}$}

\newcommand{\dphi}{$\Delta\phi$ }

\newcommand{\pt}{$p_{t}$}

\newcommand{\ptjet}{$p_{t,jet}^{rec}$}
\newcommand{\ptcut}{$p_{t}^{cut}$}
\newcommand{\ptseed}{$p_{t}^{seed}$}
\newcommand{\Rc}{$R_C$}

\begin{document}
\title{First fragmentation function measurements from full jet reconstruction in heavy-ion collisions at \rts$=200$ GeV by STAR}
\author{J\"orn Putschke\inst{1}\footnote{\email{joern.putschke@yale.edu}} (for the STAR Collaboration)}% etc
\institute{Yale University, Physics Department, New Haven, CT, US}
\date{Received: date / Revised version: date}

\abstract{
Measurements of inclusive hadron suppression and di-hadron azimuthal
correlations in ultra-relativistic nuclear collisions at RHIC have provided
important insights into jet quenching in hot QCD matter, but are
limited in their sensitivity due to well-known biases. Full jet
reconstruction in heavy-ion collisions would conceptually provide a direct
measurement of the energy of the scattered parton before energy loss,
alleviating such biases and allowing a measurement of the energy loss
probability distribution in a model-independent way from hard probes. In these proceedings  we
utilize recent progress in the reconstruction of jets in the heavy
ion environment and present the first measurement of the
fragmentation function from fully reconstructed jets in heavy
ion collisions. The fragmentation function measured in central Au+Au
collisions at \rts$=200$ GeV will be presented and discussed with respect to p+p reference measurements.
%
%\PACS{
%      {PACS-key}{discribing text of that key}   \and
 %     {PACS-key}{discribing text of that key}
 %    } % end of PACS codes
%} %end of abstract
%
}
\maketitle
\section{Introduction}
\label{intro}
Measurements of inclusive hadron suppression and di-hadron azimuthal
correlations in ultra-relativistic nuclear collisions at RHIC have provided
important insights into the properties of hot QCD matter \cite{star1,star1a}.
In particular, the high transverse momentum (high
\pt) suppression \cite{star_highpTcoor} and low \pt\ enhancement
\cite{star2} of the correlated yield of hadrons recoiling from a high
\pt\ particle (azimuthal pair separation \dphi$\sim$ $\pi$) suggest a
dramatic softening of jet fragmentation in dense matter, arising from
strong partonic energy loss. However, these measurements are limited
in their sensitivity due to well-know geometric biases (see for example \cite{Renk_Dihadron}).
%Full jet reconstruction in heavy-ion collisions would conceptually provide a direct
%measurement of the energy of the scattered parton before energy loss,
%alleviating such biases and allowing a measurement of the energy loss
%probability distribution necessary to extract properties of the
%medium in a model-independent way from hard probes. 
Full jet reconstruction in heavy-ion collisions would conceptually provide a direct
measurement of the energy of the scattered parton before energy loss,
alleviating such biases and allowing to reconstruct the partonic kinematics in an unbiased way,
independent of the fragmentation details (quenched or unquenched). This would result in a direct
measurement of the energy loss probability distribution necessary in a model-independent way.\\

A sensitive new observable accessible via full jet reconstruction is the fragmentation function
and it's modification in the presence of jet-quenching. The fragmentation function expressed in the
logarithmic variable $\xi=\ln(1/x)=\ln(E_{Jet}/p^h)$ with $x=p^h/E_{Jet}$, commonly known as the ``hump-backed'' plateau, is well-described by QCD-based MLLA (modified leading-log approximation) calculation in $e^+e^-$ collisions \cite{humpback}. Due to current experimental constraints (see sec.\ \ref{jetreco}) we will express the $\xi$ fragmentation function using
\begin{equation}
\label{eq:xi}
\xi=\ln(p_{t,jet}^{rec}/p_{t}^{hadron}).
\end{equation}
Jet quenching in the MLLA formalism has been introduced via modification of the radiation vertex coupling strength \cite{humpback}. In these calculations a significant enhancement in the fragmentation function at high $\xi$ (low hadron \pt) as well as a suppression at low $\xi$ (high hadron \pt) with respect to p+p (or $e^+e^-$) reference measurements is expected.\\

In these proceedings we will report the first fragmentation function measurement from full jet reconstruction in central Au+Au
collisions at \rts$=200$ GeV. The heavy-ion fragmentation function will be compared and discussed with respect to p+p reference measurements. In addition we will discuss the effect of the underlying high multiplicity heavy-ion background on the jet energy determination as well as  appropriate background subtraction methods in heavy-ion collisions to measure the jet energy and the fragmentation function. See \cite{sevil} for the accompanying discussion about inclusive jet spectra and systematic studies by utilizing different jet reconstruction algorithms in heavy-ion collisions, also presented for the first time during the Hard Probes 2008 meeting.

\section{Experimental setup and jet reconstruction performance} 
\label{performance}
\subsection{Experimental setup and event selection}

% ================ Figure 1 ==================
%
\begin{figure*}[t]
\vskip 0.25cm
\includegraphics[width=0.49 \textwidth]{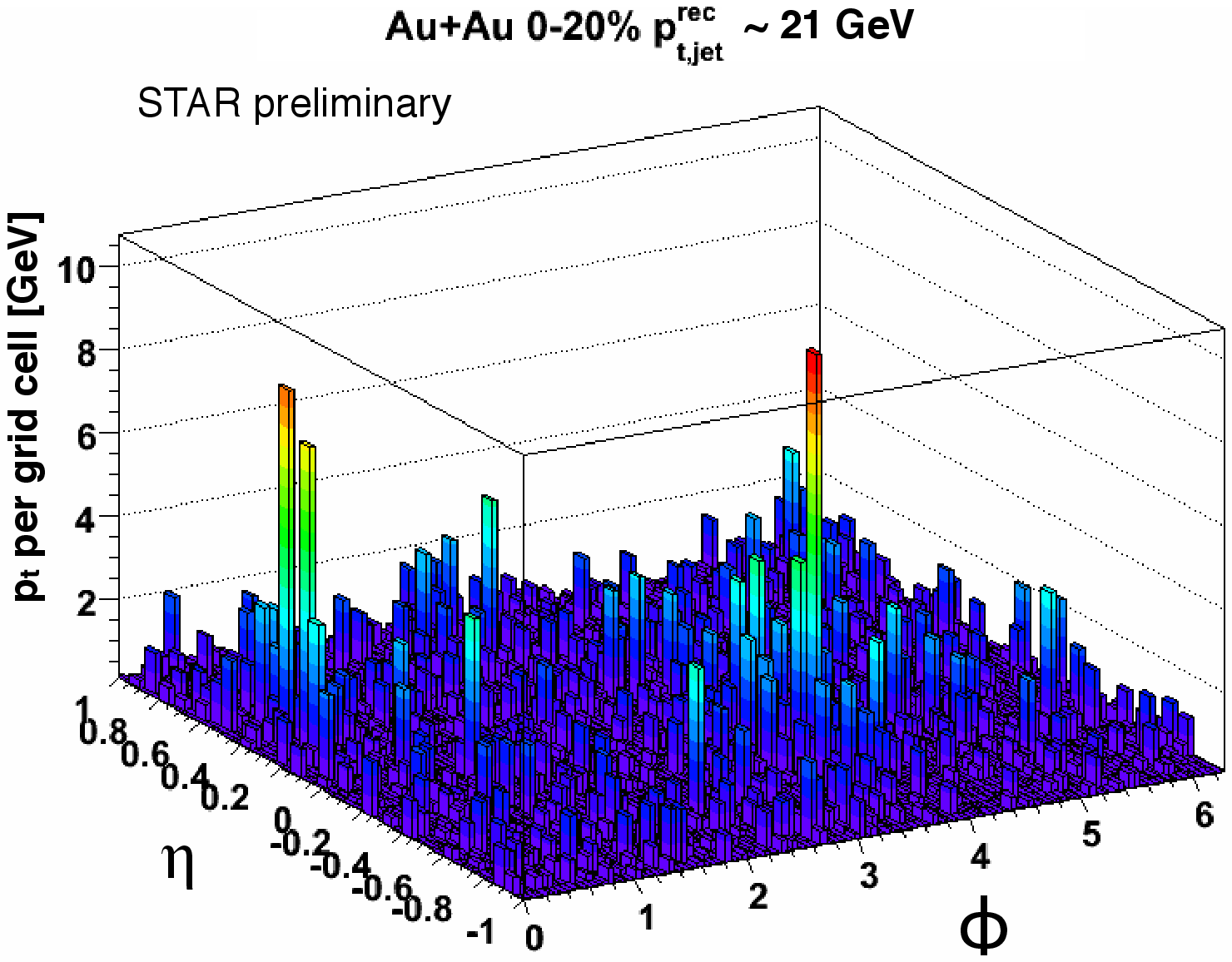}
\includegraphics[width=0.49 \textwidth]{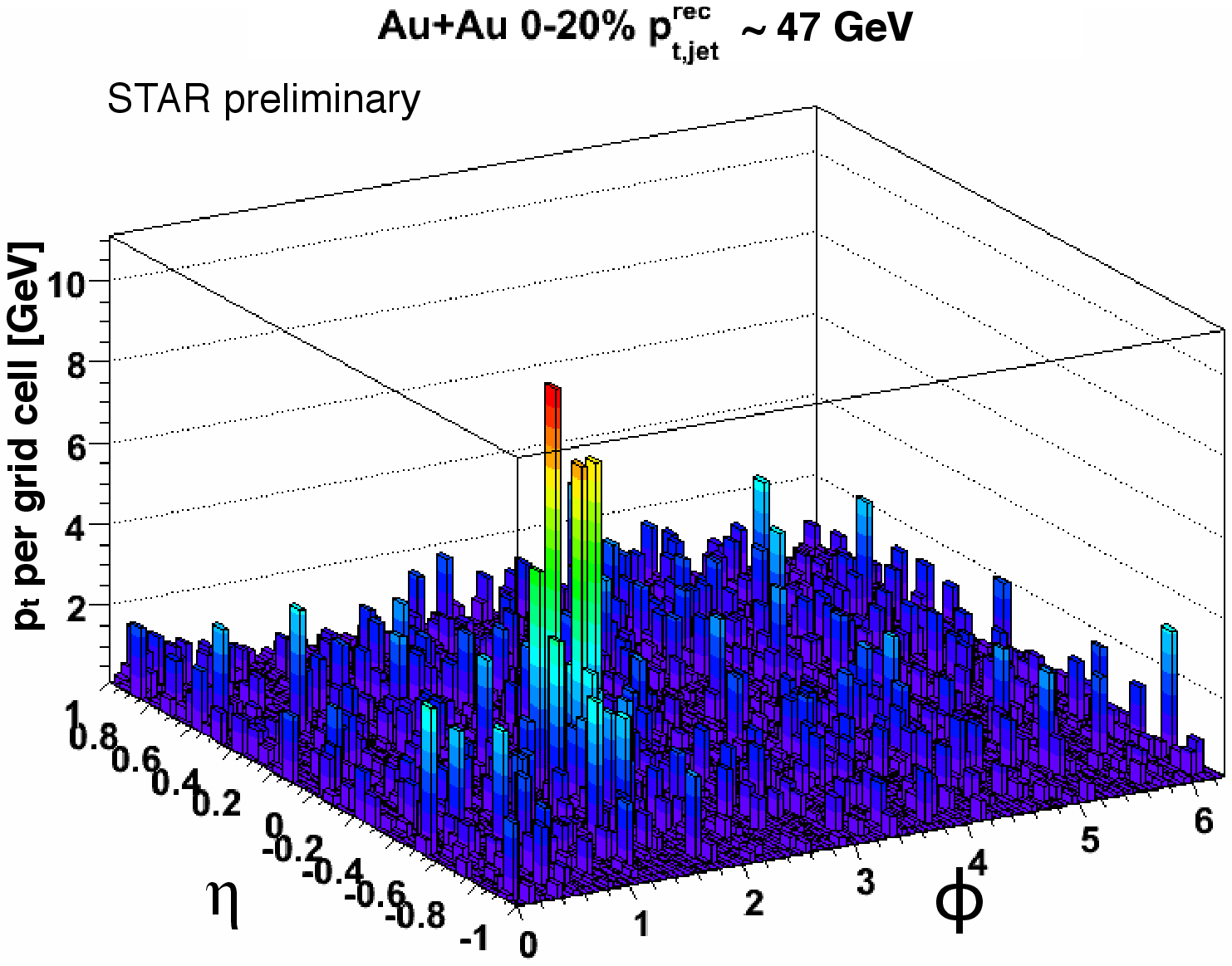}
\caption{\label{fig1} (color online)
Single event display of the summed charged and neutral \pt\ per grid cell in the $\eta$,$\phi$ plane from 0-20\% central HT triggered Au+Au collisions. Left panel: reconstructed di-jet with \ptjet$\approx$ 21 GeV. Right panel: single jet with \ptjet$\approx$ 47 GeV.} 
%\vskip -0.35cm 
\end{figure*}

% =========================================

The full jet reconstruction measurements presented in these proceedings were performed by combining the charged particles
from the STAR TPC and the neutral (tower, $0.05\times0.05$ in $\eta\times\phi$) energy from the EMCal. Both detectors, TPC and EMCal,  have full azimuthal coverage ($0 < \phi < 2\pi$) and an acceptance at mid-rapidity ranging from $-1 < \eta < 1$ (see Fig.\ \ref{fig1}). TPC primary-vertex tracks within $|\eta|<1$ were selected for this analysis using standard quality cuts for high-\pt\ measurements \cite{star1a}, which eliminate fake tracks and ensure sufficient momentum resolution. For the neutral tower energy corrections were applied to account for energy deposited by hadrons as well as to avoid double-counting of the electron energy.
For this correction electrons were identified using a $p/E$ criterium and only the \pt\ of the TPC electron track was used further on. 
In addition these identified electrons were not used in the charged particle fragmentation function measurements, because the majority 
of these are from $\pi_{0}$ conversions in the STAR detector material.\\ 
%In addition these identified electrons were not used in the charged particle fragmentation function measurements, assuming that ``all'' (neglecting the small contribution from heavy-flavor) are from $\pi_{0}$ conversions in the STAR detector material.\\ 

\noindent The following event sets were used for this analysis, based on the online trigger conditions:

\begin{itemize}
\item {\emph{ (MB Au+Au) minimum-bias trigger in Au+Au collisions}:} year 7 RHIC run coincidence condition between the two ZDCs in STAR.\\
\item {\emph{ (HT Au+Au) HT-trigger in Au+Au collisions}:} year 7 RHIC run with the trigger condition that the sum of the two highest towers in a 3x3 patch (with $E_t>5.67$ GeV for the highest tower) satisfy $E_t>7.5$ GeV (in addition to the minimum-bias trigger condition); sampling an integrated luminosity of $\sim 500\ \mu b^{-1}$ which corresponds to a p+p equivalent of $\sim 19.6\ pb^{-1}$.\\
\item {\emph{(HT p+p) HT-trigger in p+p collisions}:} year 6 RHIC run with the trigger condition for a single EMCal tower $E_t>5.4$ GeV (in addition to the minimum-bias trigger condition), corresponding to an integrated luminosity of $\sim 11\  pb^{-1}$.
\end{itemize}

\subsection{Jet reconstruction algorithms and performance}
\label{recoperf}

% ================ Figure 2 ==================
%
\begin{figure*}[t]
\includegraphics[width=0.475 \textwidth]{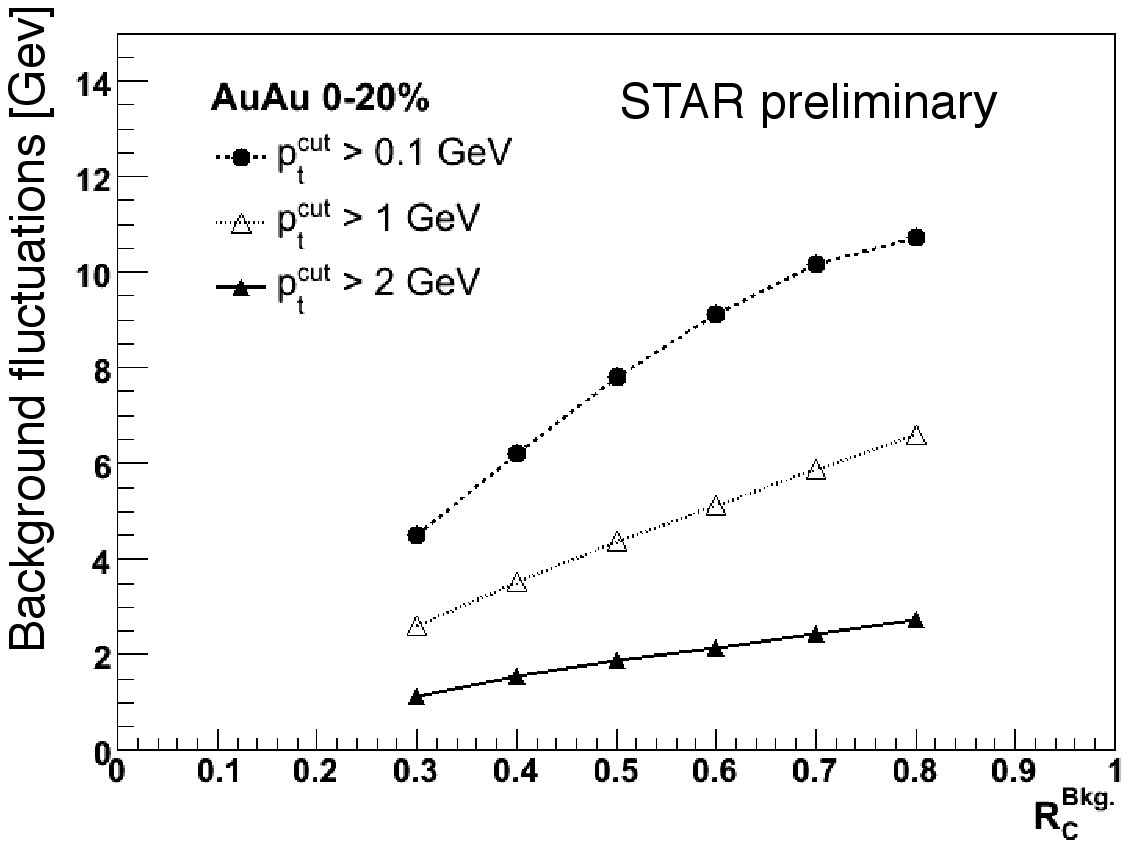}
\includegraphics[width=0.49 \textwidth]{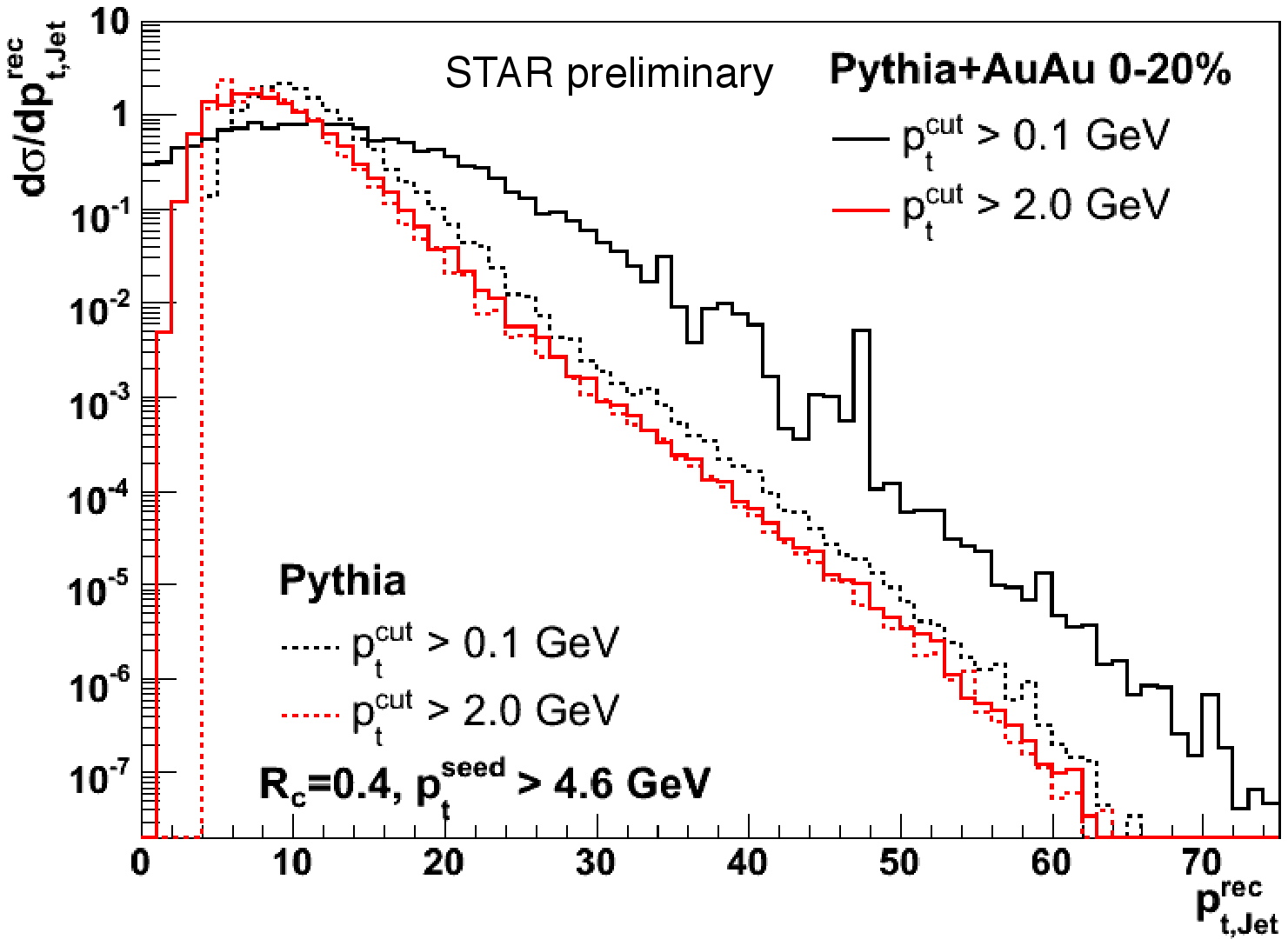}
\caption{\label{fig2} (color online) Left panel: Background fluctuations in 0-20\% central MB Au+Au collisions as function of the cone radius R for different \ptcut\ (more details see text). Right panel: Reconstructed Pythia jet spectrum with and without embedding  in 0-20\% central MB Au+Au collisions.} % for two \ptcut.} 
%\vskip -0.35cm 
\end{figure*}

% =========================================

\subsubsection{Jet reconstruction algorithms}
\label{jetreco}
If not otherwise mentioned jet-finding was performed using a seeded cone algorithm which combines all particles in the $\eta\times\phi$ plane within a cone radius of $R_C=\sqrt{\Delta\phi^2+\Delta\eta^2}$ around a high energy seed. In the presence of heavy-ion background splitting/merging and the iteration procedure to define stable cones can lead to ambiguities. As a first step we utilized a cone algorithm without splitting/merging and iteration: leading order (high seed) cone (LOCone)\footnote{$\mathtt{http://rhig.physics.yale.edu/\sim putschke/Ahijf/}$}. The relevant parameters for this algorithm are \ptseed\ and $R_C$. If not otherwise mentioned a \ptseed$=4.6$ GeV and \Rc$=0.4$ was used for all measurements. In addition a conceptually different algorithm, based on sequential recombination, was used. We choose the $k_t$ algorithm, part of the FastJet package (for more details see  \cite{fastjet} and references therein). For the LOCone a grid, based on the EMCal tower granularity ($0.05\times0.05$ in $\eta\times\phi$), was filled with charged particles and the neutral tower energy. Jet finding was performed on that grid with the summed \pt\ per grid cell\footnote{The \pt\ from charged particles was used (no PID) and $E_t=p_t$ for the neutral tower energy, assuming that a photon deposited the tower energy.}. See Fig.\ \ref{fig1} for two 0-20\% central HT triggered Au+Au events containing jets reconstructed using the LOCone. \\

In these proceedings only the jet with the highest reconstructed jet energy per event (in simulations and data) was used for jet spectra and fragmentation function measurements.

\subsubsection{Heavy-ion background and effect on jet spectrum}
\label{bkg}

% ================ Figure 3 ==================
%
\begin{figure*}[t]
\includegraphics[width=0.49 \textwidth]{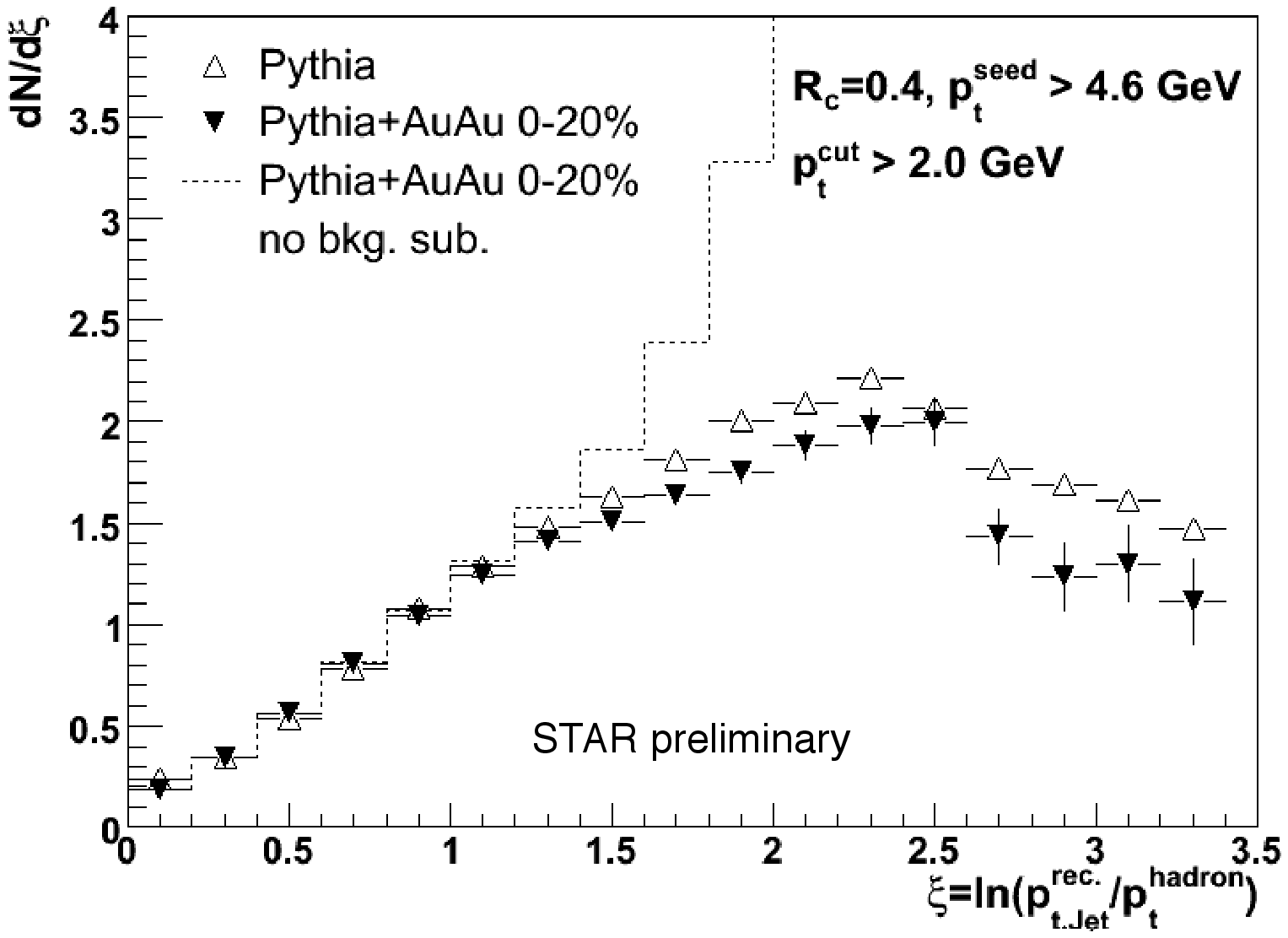}
\includegraphics[width=0.49 \textwidth]{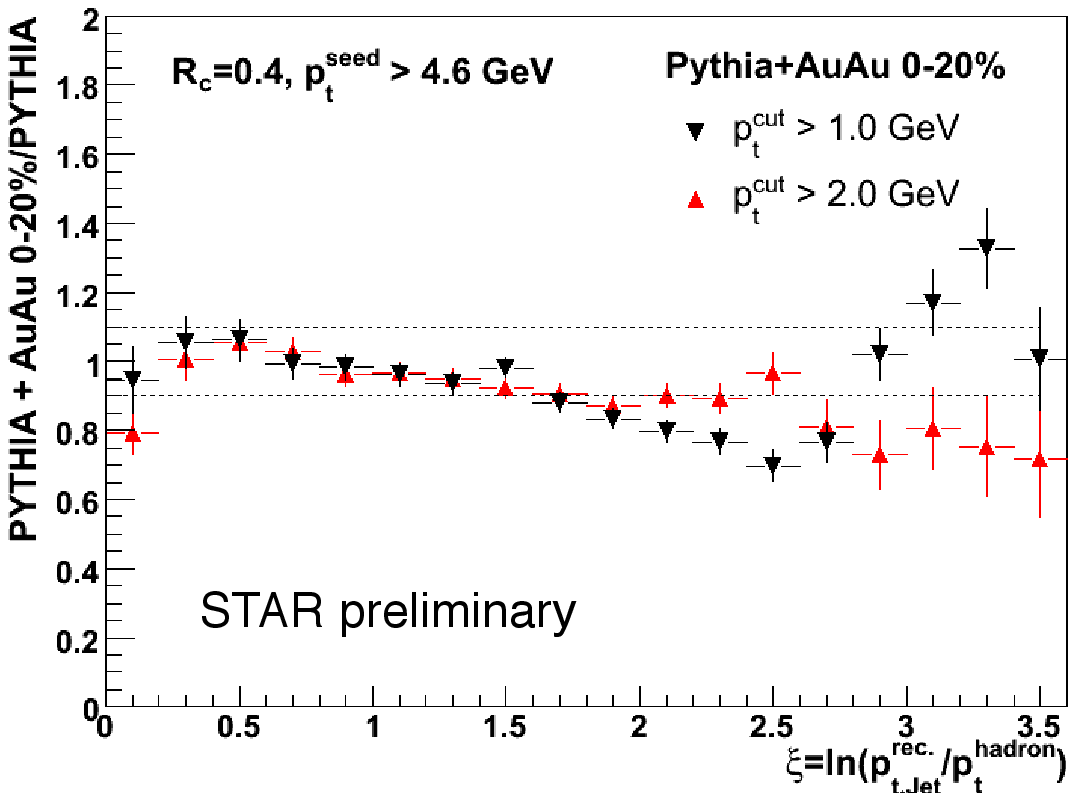}
\caption{\label{fig3} (color online)
Left panel: $\xi$ fragmentation function for 30 GeV Pythia jets embedded in 0-20\% central MB Au+Au collisions. The red dashed line represents the background contribution (for more details see text). Right panel:  $\xi$ fragmentation function for 30 GeV Pythia jets embedded in 0-20\% central MB Au+Au collisions after background subtraction compared to the Pythia fragmentation function (for reference the non-background subtracted fragmentation function is show as the dashed histogram).} 
%\vskip -0.35cm 
\end{figure*}

The main complication in performing full jet reconstruction in heavy-ion collision is the presence of the large underlying background. The common approach is to reduce the size of the cone radius \Rc\ to $0.3-0.4$ instead of $0.7$ as normally used in p+p collisions. This reduction will still recover $\sim 80$\% of the jet energy within \Rc\ $\sim 0.3$ for 50 GeV jets as observed in p+p collisions at the Tevatron \cite{tevatron}. Jet quenching in heavy-ion collision is expected to broaden the jet fragmentation and would reduce the fraction of the measured jet energy at a given cone size. This effect needs further experimental studies and theoretical exploration. To further reduce the heavy-ion background a minimum \pt\ cut (\ptcut) of up to  2 GeV on charged particles and neutral towers is applied \cite{sarah,ect}. The remaining mean background \pt\ ($<p_t^{Bkg.}>$) is estimated in the out-of-cone area, excluding the area occupied by the two highest reconstructed jets per event\footnote{Strong elliptic flow $v_2$ can lead to systematic effects in determining $<p_t^{Bkg.}>$ depending on the relative location of the jet-axis with respect to the reaction plane. With sufficient acceptance and focusing on central Au+Au collisions, where $v_2$ is small, the effect is expected to be small.}, and corrected for by scaling to the proper jet cone area. 

To study in more detail how the underlying heavy-ion background correction depends on \Rc\ and \ptcut\ we measured the region-to-region background fluctuations in 0-20\% central MB Au+Au events. The background fluctuation measurement is characterized by the gaussian width: $\sigma(\Delta p_t(R_C,p_t^{cut}))=p_t^{random}(R_C,p_t^{cut}))-<p_t^{Bkg.}>$. In Fig.\ \ref{fig2} left panel the background fluctuation for different \ptcut\ is shown as function of \Rc. One can see that applying a \ptcut\ of 2 GeV significantly reduces the background fluctuation from $\sim$ 6 GeV for  \Rc\ $=0.4$ with \ptcut\ $=0.1$ GeV to below 2 GeV for \ptcut\ $=2$ GeV . 

The effect of background fluctuations on the jet energy resolution and on the reconstructed jet spectrum was studied by embedding a Pythia generated jet-spectrum \cite{Pythia8} (\emph{Pythia}) into 0-20\% central MB Au+Au collisions (\emph{Pythia+AuAu 0-20\%}). 
For the jet signal the charged and electromagnetic particles were taken directly form Pythia\footnote{The detector effects are expected to be small especially compared to the uncertainties caused by the heavy-ion background.}.
%Only the charged and neutral Pythia particle  information in the experimental acceptance was used, no further detector simulation response was applied\footnote{The detector effects are expected to be small especially compared to the uncertainties caused by the heavy-ion background.}.  
In Fig.\ \ref{fig2} right panel the reconstructed Pythia jet spectrum with and without embedding in 0-20\% central MB Au+Au collisions is shown for a \ptcut$=0.1$ and $2$ GeV. The jet energy resolution in heavy-ion collisions is mainly dominated by the background fluctuations (see \cite{sevil} for more details). It is apparent that for a \ptcut$=0.1$ GeV the steeply falling jet spectrum folded with the jet energy resolution $-$ determined by the background fluctuations $-$ leads to a significant broadening of the reconstructed jet spectrum in Au+Au collisions compared to the input Pythia spectrum. As a consequence, for a \ptcut$=0.1$ GeV, jets with a reconstructed jet energy of \ptjet$>30$ GeV in the presence of background will be dominated by the feed-down of lower \ptjet\ jets. In contrast the jet spectra  for a \ptcut$=2$ GeV with and without embedding in 0-20\% central MB Au+Au collisions are in very good agreement, therefore allowing a good determination of \ptjet\ on an event-by-event basis in Pythia+AuAu 0-20\% as compared to Pythia. %(representing the p+p case).

\subsubsection{Fragmentation function measurement performance}
\label{xisim}

To measure the $\xi$ fragmentation function the jet energy \ptjet\ has to be determined on an event-by-event basis. To achieve this, one has to apply a \ptcut\ of 2 GeV to suppress the effect of background fluctuations, as discussed in sec.\ \ref{bkg}. In the following the  fragmentation function measurements were performed using the LOCone with \Rc$=0.4$ and the $k_t$ algorithm with $R_{Param}=0.4$ and a \ptcut$=2$ GeV to determine the jet energy \ptjet. In contrast, all charged particles in a cone of $R_C^{FF}=0.7$ (around the jet axis) without applying an additional \pt\ were included in the fragmentation function measurements. In Fig.\ \ref{fig3} left panel the charged $\xi$ fragmentation function of a 30 GeV Pythia jet embedded in 0-20\% central MB Au+Au collisions is shown. For $\xi$ values larger then two the fragmentation function is dominated by the soft underlying heavy-ion background. The red dashed curve represents the $\xi$ background estimate determined in the out-of-cone area (see sec.\ \ref{bkg}) on an event-by-event basis after scaling to \ptjet\ and the jet fragmentation function area. Fig.\ \ref{fig3} right panel shows the comparison of the the background corrected $\xi$ fragmentation function in Pythia+AuAu 0-20\% to Pythia. The used background subtraction method and the jet energy resolution causes deviations of less then $10-20$\% for $\xi<2.5$ assuming Pythia fragmentation. The deviations in the $\xi$ shape for Pythia+AuAu 0-20\% can be largely explained by jet energy resolution effects. For $\xi>2.5-3$ the fragmentation function measurement is dominated by the underlying heavy-ion background. Without further studies we will exclude these higher $\xi$ regions in the following discussion.
\begin{figure*}[t]
\begin{center}
\includegraphics[width=0.49 \textwidth]{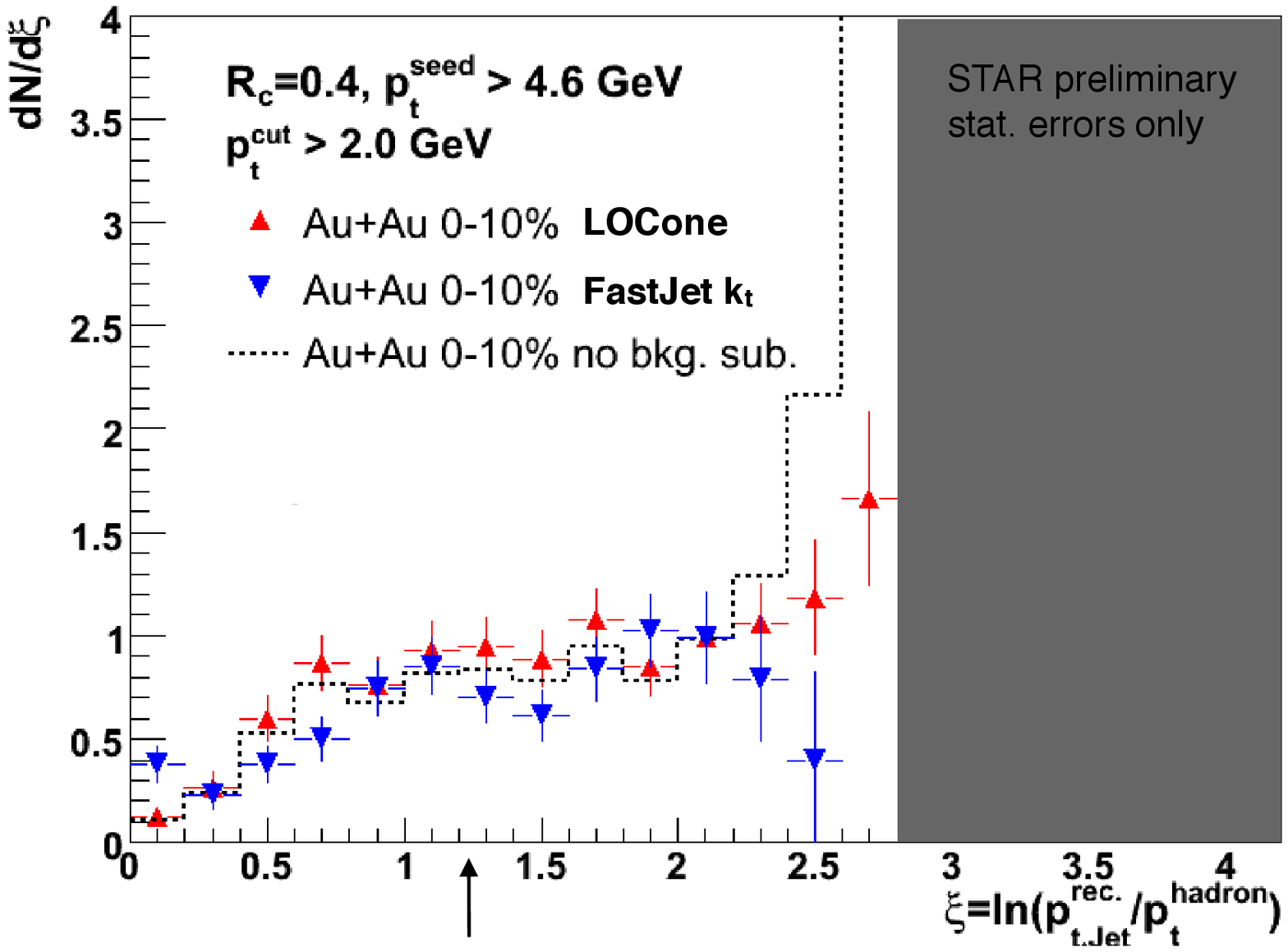}
\includegraphics[width=0.49 \textwidth]{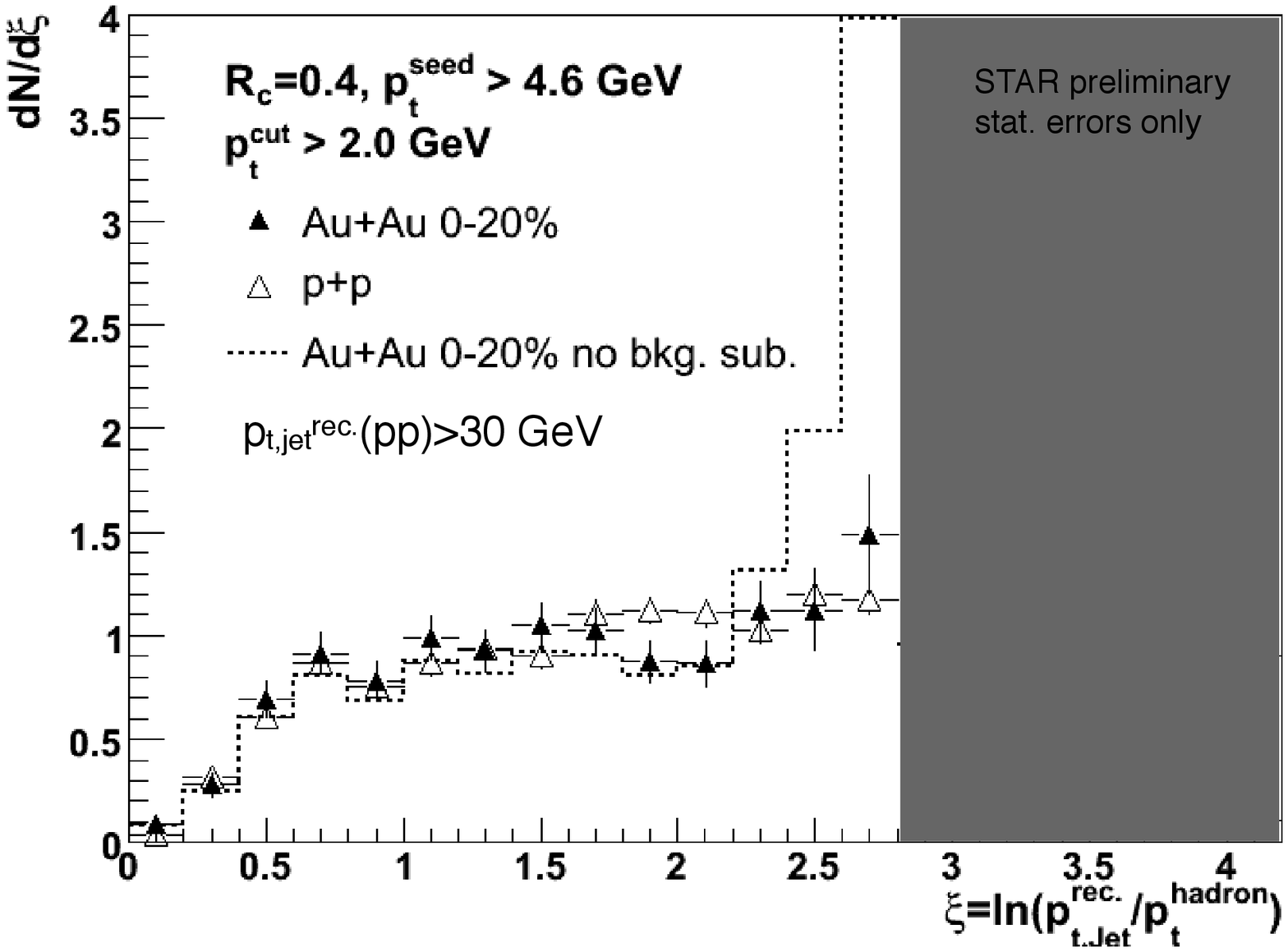}
%\vskip -0.35cm 
\end{center}
\caption{\label{fig4} (color online) Left panel: $\xi$ fragmentation function in 0-10\% central HT Au+Au collisions after background subtraction applying a \ptcut$> 2$ GeV for \ptjet$>30$ GeV using the LOCone and FastJet $k_t$ jet-finding algorithm. The arrow indicates the $\xi$ value for a hadron with \pt$\approx10$ GeV. Right panel:  $\xi$ fragmentation function in 0-20\% central HT Au+Au collisions compared to HT p+p collisions for \ptjet$(pp)>30$ GeV using the LOCone with \Rc$=0.4$ and a \ptcut$> 2$ GeV (for reference the non-background subtracted fragmentation function is show as the dashed histogram). The grey shaded area in both panels indicates the region in $\xi$ (corresponding to $p_t^{hadron}<2$ GeV) where the Au+Au measurement is expected to be dominated by the underlying background (see discussion in sec.\ \ref{xisim}).} 

\end{figure*}
% =========================================

% =========================================

\section{Fragmentation function measurements in Au+Au (and p+p) collisions}
\label{ff}

% ================ Figure 5 ==================
%
\begin{figure*}[t]
\begin{center}
\includegraphics[width=0.65 \textwidth]{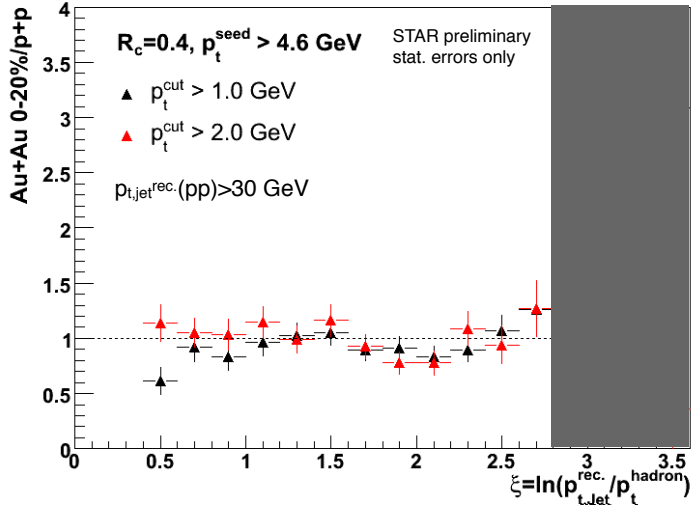}
%\vskip -0.35cm 
\end{center}
\caption{\label{fig5} (color online)
Ratio of  the $\xi$ fragmentation function measured in 0-20\% central HT Au+Au collisions to HT p+p collisions for \ptjet$(pp)>30$ GeV with \ptcut$>1$ and $2$ GeV (more details see text). The grey shaded area indicates the region in $\xi$  (corresponding to $p_t^{hadron}<2$ GeV) where the Au+Au measurement is expected to be dominated by the underlying background (see discussion in sec.\ \ref{xisim}).} 

\end{figure*}
% =========================================

Using the algorithms and background correction procedure as discussed in sec.\ \ref{recoperf}, we performed the first fragmentation function measurement from full jet reconstruction in heavy-ion collisions using central HT Au+Au collisions at \rts$=200$ GeV. Fig.\ \ref{fig4} left panel shows the  $\xi$ fragmentation function for \ptjet$>30$ GeV in 0-10\% central HT Au+Au collisions after background subtraction using the LOCone and FastJet $k_t$ jet-finding algorithm. A  \ptcut$> 2$ GeV was applied and for the LOCone a \Rc$=0.4$ and for the FastJet $k_t$ a $R_{Param}=0.4$ was used to suppress the effect of background fluctuations (see discussion in sec.\ \ref{xisim}). The comparison of the measured fragmentation function in 0-10\% central HT Au+Au collisions for $\xi<2.5$ by utilizing two conceptually different jet finding algorithm, LOCone and  FastJet $k_t$, given the applied parameters, is a first step to address systematic uncertainties in full jet reconstruction in heavy-ion collisions. Further systematic studies are needed, but the agreement of LOCone and FastJet $k_t$ gives us confidence in our jet reconstruction performed on central HT Au+Au collisions.

In Fig.\ \ref{fig4} right panel the $\xi$ fragmentation function in 0-20\% central HT Au+Au collisions compared to HT p+p collisions for \ptjet$(pp)>30$ GeV using the LOCone with \Rc$=0.4$ and a \ptcut$> 2$ GeV is shown. In order to select a corresponding p+p jet population in  0-20\% central HT Au+Au collisions Pythia simulations (see sec.\ \ref{bkg}) were used to determine the appropriate \ptjet$(AuAu)$ region: \ptjet$(AuAu)>31$ GeV for \ptcut$>2$ GeV and \ptjet$(AuAu)>35$ GeV for \ptcut$>1$ GeV.

To quantify in more detail the modifications in the fragmentation function due to quenching effects in Au+Au with respect to p+p reference measurements, the $\xi$ fragmentation function ratio of Au+Au to p+p for \ptjet$(pp)>30$ GeV is shown in Fig.\ \ref{fig5}. %applying a \ptcut$>2$ and 1 GeV. 
No apparent modification in the  Au+Au $\xi$ fragmentation function is seen in the $\xi$ range\footnote{The lower $\xi$ cut off is currently motivated by the momentum resolution of charged tracks above \pt$>10$ GeV. Further studies are needed to estimate and correct for the momentum resolution effects at highest \pt.} $0.5<\xi<2.75$ for \ptcut$>2$ GeV (and 1 GeV). 

One should note that the presented $\xi$ distributions are not corrected for the missing jet energy imposed by applying a \ptcut\ (and \Rc) cut. This fraction could be estimated assuming Pythia fragmentation (see sec.\ \ref{bkg}), but is expected to be different due to quenching effects in Au+Au collisions (more discussion in sec.\ \ref{discussion}). 

\section{Interpretation and discussion}
\label{discussion}

% ================ Figure 6 ==================
%
\begin{figure*}[t]
\begin{center}
\includegraphics[width=0.65 \textwidth]{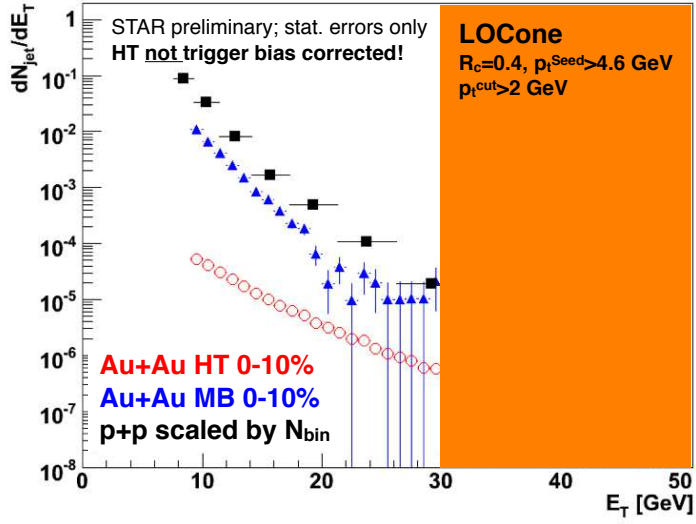}
%\vskip -0.35cm 
\end{center}
\caption{\label{fig6} (color online) Inclusive jet spectrum for 0-10\% MB and HT Au+Au collisions compared to \nbin\ scaled p+p reference measurements \cite{spin}. The MB Au+Au inclusive jet spectrum is correct for missing energy caused by applying a \ptcut\ and \Rc\ cut as well as for efficiency using Pythia simulations (for more details see text and \cite{sevil}). The HT Au+Au inclusive jet spectrum is not corrected. The orange shaded area indicates the \ptjet\ region in which the $\xi$ fragmentation function measurements were performed (see sec.\ \ref{ff}). } 
\end{figure*}
% =========================================

\begin{figure*}[t]
\begin{center}
\includegraphics[width=\textwidth]{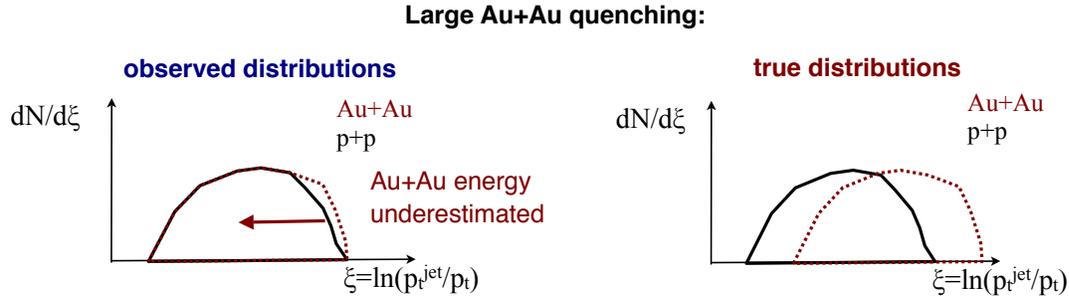}
%\vskip -0.35cm 
\end{center}
\caption{\label{sketch} (color online)
Sketch to illustrate the effect of jet-energy bias due to energy loss and the applied \ptcut\ on the measured $\xi$ fragmentation function.} 

\end{figure*}

Assuming that full jet reconstruction allows us to select an unbiased jet sample (\raajet$=1$) and with the measured inclusive hadron suppression in heavy-ion collisions (\raa\ $<1$) one would expect to the see modifications in the fragmentation function measurement due to jet quenching in heavy-ion collisions with respect to p+p reference measurements. 
In this light, our finding that there seems to be no modification in the fragmentation function in Au+Au collisions (see sec.\ \ref{ff}) is very surprising. To investigate further whether an unbiased sample of jets is selected in the fragmentation function analysis, we compare the jet spectra for 0-10\% MB  (corrected for \ptcut, \Rc\ and efficiency using Pythia simulations) and HT Au+Au collisions (no correction applied) to the \nbin\ scaled p+p reference measurements \cite{spin} as shown in Fig.\ \ref{fig6}. 
Using a \ptcut of 2 GeV in the MB Au+Au jet spectrum measurement suppresses the jet yields with respect to the \nbin\ scaled p+p jet spectrum. Applying a \ptcut\ in heavy-ion collisions leads to \raajet$<1$, indicating a bias in the selected jet population in Au+Au collisions. However, measurements of the inclusive jet spectrum without applying an additional \ptcut\ indicate that we might be able to recover an unbiased jet sample (\raajet$\approx1$)  (see \cite{sevil}). This result suggests that in the presence of jet quenching in heavy-ion collision the correction for the \ptcut\ and reduced \Rc\  based on Pythia is not sufficient and does seem to underestimate the reconstructed jet energy in heavy-ion collisions. To illustrate this effect on the fragmentation function measurements see the sketch in Fig.\ \ref{sketch}. In the presence of large quenching correcting the jet energy based on unmodified jet fragmentation from Pythia will underestimate the reconstructed jet energy in central Au+Au collisions. The true jet energy in Au+Au is higher than in p+p collisions and will therefore lead to a relative shift of the $\xi$ fragmentation function in Au+Au with respect to p+p, as illustrated in Fig.\ \ref{sketch} right panel. This relative shift would then lead to the expected modification pattern of the fragmentation function in Au+Au due to jet quenching (see sec.\ \ref{intro}).

An alternative scenario that would describe the data is the ``black-and-white picture", where some partons lose so much energy that they are ``lost"  (\raajet$<1$), while the remaining fraction of jets suffers little or no modification, so that the fragmentation function of the ``surviving" jets is unmodified. To distinguish between these two scenarios one has to establish if \raajet\ using our jet reconstruction framework is unity (or below).\\

In addition the HT trigger requirement in p+p and Au+Au will lead to a \emph{trigger bias} in the jet spectra and fragmentation functions. Recent studies in p+p collisions at \rts$=200$ GeV \cite{elena} indicate that for a HT trigger with a tower energy threshold greater than $5.4$ GeV biases in the fragmentation function are still present at \ptjet$(pp)\sim30$ GeV. To study the HT trigger bias in the presence of jet quenching effect, by even applying a higher trigger threshold of $7.5$ GeV, further statistics in the MB Au+Au data sample is needed (and available on tape) to assess the bias at higher \ptjet. This trigger bias could lead $-$ in the \ptjet\ range of $\gtrsim 30$ GeV used for the fragmentation function measurements $-$ to a bias in the selected jet population favoring jets with no or only small energy loss. 

% ================ Figure 7 ==================

% =========================================

\section{Summary and Outlook}

In these proceedings the first fragmentation function measurement from full jet reconstruction in heavy-ion collisions was presented (see sec.\ \ref{ff}). It was shown that using a smaller cone radius of \Rc$=0.4$ and applying a \ptcut of $2$ GeV reduces significantly the effect of the underlying background fluctuations in our jet measurements, allowing $-$ based on Pythia simulations (see sec.\ \ref{bkg}) $-$ a reasonably precise determination of \ptjet\ on an event-by-event basis necessary for the measurement of the fragmentation function. Utilizing the current jet finding algorithms and background subtraction procedures deviations of less then $10-20$\% for $\xi<2.5$ assuming Pythia fragmentation are observed in the $\xi$ fragmentation function measurement (see sec.\ \ref{xisim}).
However, applying a \ptcut to suppress the heavy-ion background leads to a bias in the selected jet-population which would result in an  underestimation of the true jet energy \ptjet$(AuAu)$ in the presence of large jet quenching effects (see discussion in sec.\ \ref{discussion}). This could explain why in our measurement we do not see an apparent  modification in the fragmentation function in Au+Au collisions. To correct properly for this effect  further experimental and theoretical developments are needed. First calculations addressing the effect of experimental cuts on the reconstructed jet energy in the presence of quenching can be found in \cite{Renk_Jet}.\\

The possible additional bias in our used data-sets caused by the online trigger requirement can be assessed in the near future by analyzing the full available year 7 RHIC run MB Au+Au statistics recorded by STAR (see also \cite{sevil}).

\begin{figure*}[t]
\begin{center}
\includegraphics[width=0.60 \textwidth]{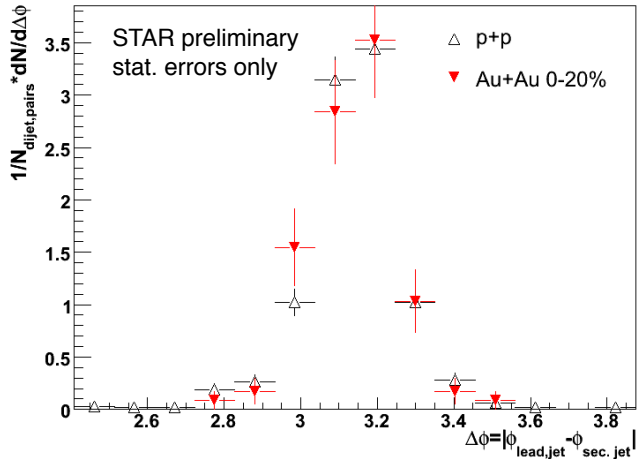}
%\vskip -0.35cm 
\end{center}
\caption{\label{fig8} (color online) Di-jet $\Delta\phi$ distribution for \ptjet$>20$ GeV in 0-20\% central HT Au+Au and HT p+p collisions using the LOCone algorithm with \Rc$=0.4$, \ptcut$=2$ GeV and \ptseed$=4.6$ GeV. } 
\end{figure*}

Another possibility to study the trigger bias and to select a trigger bias free jet population utilizing the current HT Au+Au data-set is to select the (unbiased) recoil jet. In Fig.\ \ref{fig8} the di-jet $\Delta\phi$ distribution for \ptjet$>20$ GeV in 0-20\% central HT Au+Au and HT p+p collisions using the LOCone algorithm with \Rc$=0.4$, \ptcut$=2$ GeV and \ptseed$=4.6$ GeV is shown.  Systematic studies are needed to utilize the di-jet pairs for further physics analysis. But given the current di-jet statistics it might be possible to extract a trigger bias free fragmentation function in Au+Au collisions. In addition di-jets in p+p collisions can be used to estimate the jet energy resolution in a model independent way (see for example \cite{elena}). \\

In the future further systematic studies utilizing the currently available jet finding algorithms are needed, and will be carried out, to address systematic and conceptual uncertainties from full jet reconstruction in heavy-ion collisions.

% ================ Figure 8 ==================

% =========================================

% % ========= References ========= 
\bibliographystyle{epj}
\bibliography{ref}

\end{document}